\documentclass[onecollarge,natbib]{svjour2}
\bibpunct{[}{]}{;}{n}{}{,} 
\smartqed  
\usepackage{graphicx}
\usepackage{amsmath}

\def\tstrut{\vrule height2.5ex depth0pt width0pt} 

\journalname{Few-Body Systems (EFB22)}

\begin{document}

\title{Semileptonic and nonleptonic decays of $B_s$ mesons
}


\author{C. Albertus
}


\institute{C. Albertus \at
              Departamento de F\'\i sica At\'omica, Molecular y Nuclear \\
              Facultad de Ciencias. Universidad de Granada.
              Avenida de Fuentenueva S/N 37008 Granada Spain
              \email{albertus@ugr.es}           
}

\date{Received: date / Accepted: date}

\maketitle

\begin{abstract}

We investigate the semileptonic and nonleptonic decays of $B_s$
mesons. We work within the context of nonrelativistic constituent
quark models. We calculate the different form factors that parameterize
the hadron matrix elements.

\end{abstract}


\vspace{12pt} In this work we study the semileptonic decays of $B_s$
mesons into $D_s$ states and some of its nonleptonic decays. We work
in the nonrelativistic constituent quark model scheme. Since the
discovery of the $B_s$ and $\bar B_s$ mesons, their lifetimes and
decay modes have been in the aim of both theoretical and experimental
collaboration and, for instance, are currently being measured in the
LHCb experiment at CERN
\cite{Aaij:2013pua,Aaij:2013iua,Guadagnoli:2013mru,Blusk:2012it,Dewhurst:2011zz,Aad:2012kba,Giurgiu:2010zz,Abulencia:2005ia,Abe:1995ka}.

The description of the different $J^P$ states involved is given in
Ref.~\cite{albertus:prep} and is omitted here for the shake of
brevity. The semileptonic decays are governed by the V-A $b\to c$
current. The matrix elements are parame\-trized in terms of form
factors as in Eq.~(9) of Ref.~\cite{albertus:prep}
%
The calculation of
the matrix elements and form factors can be found in full detail in
Ref.~\cite{albertus:prep}, as well.

The semileptonic double differential decay width with respect to $q^2$
and the cosine, $x_l$, of the angle between the final meson momentum
and the momentum of the final charged lepton, the latter measured in
the lepton-neutrino center of mass frame (CMF), results to be, for a
$\bar B_s$ at rest
\begin{align}
\frac{d^2\Gamma}{dx_l dq^2}=&\frac{G_F^2}{64m^2_{\bar B_s}}\frac{|V_{bc}|^2}{8\pi^3}\frac{\lambda^{1/2}(q^2,m^2_{\bar B_s},m^2_{c\bar s})}{2m_{\bar B_s}}
\frac{q^2-m_l^2}{q^2}
{\cal H}_{\alpha\beta}(P_{\bar B_s},P_{c\bar s}){\cal L}^{\alpha\beta}(p_l,p_\nu),
\label{eq:diffgamma}
\end{align}
where $G_F$ \cite{PhysRevD.86.010001} is the Fermi constant,
$\lambda(a,b,c)=(a+b-c)^2-4ab$, $m_l$ is the mass of the charged
lepton, ${\cal H}$ and ${\cal L}$ are the hadron and lepton tensors,
and $p_l$ and $p_\nu$ are the lepton four momenta. $P_{\bar B_s}$ and
$P_{c\bar s}$ (with $c\bar s =D_s, D_s^*, D_{s2}^*$) are the meson
four-momenta, $m_{\bar B_s}$ and $m_{c\bar s}$ their masses
respectively, $P=P_{\bar B_s} + P_{c\bar5 s}$, and $q=P_{\bar B_s} -
P_{c\bar s}$. To calculate the contraction of the lepton and hadron
tensors, that involve the form factors from Eq. (9) of \cite{albertus:prep}, we
shall follow the helicity formalism of Ref.~\cite{Ivanov:2005fd}. This
calculation is given in full detail in Sec. III E of
Ref.~\cite{albertus:prep}. Integrating Eq.~\ref{eq:diffgamma} we
obtain the total decay widths. Numerical values can be found in
\cite{albertus:prep}.

In this work we have also studied the nonleptonic $\bar B_s \to c\bar
s M_F$ two-meson decays, where $M_F$ is a pseudoscalar or vector
meson.  These decays correspond to a $b \to c$ transition at the quark
level as well. These transitions are governed, neglecting penguin
operators, by the effective Hamiltonian
\cite{Ebert:2006nz,Beneke:1999br}
\begin{equation}
H_{\rm
  eff}=\frac{G_F}{\sqrt2}\left(V_{cb}\left[c_1(\mu)Q_1^{cb}+c_2(\mu)Q_2^{cb}\right]+H.c.\right),
\end{equation}
where $c_{1,2}$ are scale-dependent Wilson coefficients, and $Q_{1,2}$
are local four-quark operators given by
\begin{align}
Q_1^{cb}=&\bar{\Psi}_c(0)\gamma_\mu(I-\gamma_5)\Psi_b(0)
\left[V_{ud}^*\bar{\Psi}_d(0)\gamma^\mu(I-\gamma_5)\Psi_u(0)
+V_{us}^*\bar{\Psi}_s(0)\gamma^\mu(I-\gamma_5)\Psi_u(0)
\right.
\nonumber\\&
\hspace{3.2cm}+V_{cd}^*\bar{\Psi}_d(0)\gamma^\mu(I-\gamma_5)\Psi_c(0)
\left.
+V_{cs}^*\bar{\Psi}_s(0)\gamma^\mu(I-\gamma_5)\Psi_c(0)\right]
\nonumber\\
Q_2^{cb}=&\bar{\Psi}_d(0)\gamma_\mu(I-\gamma_5)\Psi_b(0)\left[
V_{ud}^*\bar{\Psi}_c(0)\gamma^\mu(I-\gamma_5)\Psi_u(0)
+V^*_{cd}\bar{\Psi}_c(0)\gamma^\mu(I-\gamma_5)\Psi_c(0)\right]\nonumber\\
+&\bar{\Psi}_s(0)\gamma_\mu(I-\gamma_5)\Psi_b(0)\left[
V_{us}^*\bar{\Psi}_c(0)\gamma^\mu(I-\gamma_5)\Psi_u(0)
+V^*_{cs}\bar{\Psi}_c(0)\gamma^\mu(I-\gamma_5)\Psi_c(0)\right],
\end{align}
where $V_{ij}$ are CKM matrix elements. We follow the factorization
approximation, i. e., the hadron matrix elements of the effective
Hamiltonian are evaluated as a product of quark-current matrix
elements. One of these is that of the $B_s$ transition to one of the
final mesons, while the other one corresponds to a transition of the
vacuum to the second final mesons, which is given by the corresponding
meson decay constant. For $M_F=\pi,\rho,K$ or $K^*$ the decay width is
given by
\begin{align}
\Gamma&=\frac{G_F^2}{16\pi m_{B_s}^2}|V_{bc}|^2|V_{F}|^2
\frac{\lambda^{1/2}(m_{B_s}^2,m_{c\bar s}^2,m_{M_F})^2}{2m_{B_s}}a_1^2
{\cal H}_{\alpha\beta}(P_{B_s},P_{c\bar s})\hat{\cal H}^{\alpha\beta}(P_{F}),
\end{align}
where $a_1\approx 1.14$ is a Wilson coefficient and $V_{F}$ is a CKM
matrix elements which depends on the actual decay considered. As
explained in Ref.~\cite{albertus:prep}, the contraction of the
resulting tensors has been performed using the helicity formalism.

Some results for the semileptonic and nonleptonic decay widths can be
found in Table~\ref{tab:comp1}. More results and comparison with
previous calculations have been published in
Ref.~\cite{albertus:prep}.

\begin{table}
\caption{\label{tab:comp1} Branching fractions for the indicated decay channels, in percentage.\label{tab:bracomp}}
\centering
\begin{center}
\begin{tabular}{c|c||c|c}
\hline\hline\tstrut
                                      & BR in \% & BR in \%\\\hline\tstrut
$\bar B_s \to D_s^+ e^- \bar \nu_e$              &  2.32   & $\bar B_s \to D_s^+ \pi^-$       & 0.53  \\
$\bar B_s \to D_s^{*+} e^-\bar \nu_e$             &  6.26   & $\bar B_s \to D_s^+ \rho^-$      & 1.26  \\
$\bar B_s \to D_s^+ \tau^- \bar \nu_\tau$         &  0.67   & $\bar B_s \to D_s^+ K^-$         & 0.04  \\
$\bar B_s \to D_s^{*+} \tau^-\bar \nu_\tau$        &  1.53   & $\bar B_s \to D_s^+ K^{*-}$       & 0.08\\\hline
\tstrut
$\bar B_s \to D_{s0}^{*+} \mu^-\bar \nu_\mu$       &0.39     &$\bar B_s \to D_{s0}^{*+} \pi^-$       & 0.10\\
$\bar B_s \to D_{s1}^{*+}(2460) \mu^-\bar \nu_\mu$ &0.60      &$\bar B_s \to D_{s0}^{*+} \rho^-$      & 0.27\\
$\bar B_s \to D_{s1}^{*+}(2536) \mu^-\bar \nu_\mu$ &0.19     &$\bar B_s \to D_{s0}^{*+} K^-$         & 0.009\\
$\bar B_s \to D_{s2}^{*+} \mu^-\bar \nu_\mu$       &0.44     &$\bar B_s \to D_{s0}^{*+} K^{*-}$       & 0.16\\
\hline
\end{tabular}
\end{center}
\end{table}

\begin{acknowledgements}
The author thanks a contract granted by the CPAN project and support
from Universidad de Granada and Junta de Andaluc\'\i a, through the
grant FQM-225.
\end{acknowledgements}

\bibliographystyle{unsrt}
\bibliography{biblist.efb22}   

%
%

\end{document}